\documentclass[12pt]{article}

\usepackage[english]{babel}
\usepackage{amsfonts}
\usepackage{amssymb}
\usepackage{graphicx}
\newcommand{\be}{\begin{equation}}
\newcommand{\ee}{\end{equation}}
\newcommand{\bea}{\begin{eqnarray}}
\newcommand{\eea}{\end{eqnarray}}

\begin{document}

\begin{titlepage}

\begin{flushright}
{\tt
    hep-th/0512179}
 \end{flushright}

\bigskip

\begin{center}

{\bf \Large{{Static quantum corrections to the Schwarzschild
spacetime}}}

\bigskip
\bigskip\bigskip
 A.  Fabbri$^a$,\footnote{afabbri@ific.uv.es} S.
Farese$^a$,\footnote{farese@ific.uv.es}
 J. Navarro-Salas$^a$,\footnote{jnavarro@ific.uv.es} G. J. Olmo$^{b}$,\footnote{olmoalba@uwm.edu}
 and H. Sanchis-Alepuz$^a$,\footnote{helios.sanchis@ific.uv.es}\footnote{Talk given by J. N-S at the Conference ``Constrained dynamics and quantum gravity, QG05'', Cala Gonone, Italy, September 2005}

\end{center}

\bigskip%

\footnotesize \noindent {\it a) Departamento de F\'{\i}sica
Te\'orica and
    IFIC, Centro Mixto Universidad de Valencia-CSIC.
    Facultad de F\'{\i}sica, Universidad de Valencia,
        Burjassot-46100, Valencia, Spain.}\\
{\it b) Department of Physics, University of Wisconsin-Milwaukee,
P.O. Box 413, Milwaukee, Wisconsin, 53201 USA}

\bigskip

\bigskip
\begin{abstract}
We study static quantum corrections of the Schwarzschild metric in
the Boulware vacuum state. Due to the absence of a complete
analytic expression for the full semiclassical Einstein equations
 we approach the problem
by considering the $s$-wave approximation and solve numerically
the associated backreaction equations. The  solution, including
quantum effects due  to pure vacuum polarization, is similar to
the classical Schwarzschild solution up to the vicinity of the
classical horizon. However, the radial function has a minimum at a
time-like surface  close to the location of the classical event
horizon. There  the $g_{00}$ component of the metric reaches a
very small but non-zero value. The analysis unravels how a
curvature singularity emerges beyond this bouncing point. We
briefly discuss the physical consequences of these results by
extrapolating them to a dynamical collapsing scenario.

\end{abstract}

\end{titlepage}

\newpage

\section{Introduction}
The most important  results on  quantum properties of black holes
\cite{hawk1, parker, wald} (see also \cite{birreldavies,
frolovnovikov, icp05}) are obtained in the so-called fixed
background approximation. This means that the spacetime background
is assumed to be fixed and not modified by the  quantum behavior
of matter. There are two reasons for doing this. First, one
expects that the inclusion of backreaction effects will not modify
essentially the physical results obtained in the fixed background
approximation, at least until reaching the Planck scale. Second,
to go beyond this approximation requires to solve the
semiclassical Einstein equations
\begin{equation} \label{Einsteinsemiclassical}
G_{\mu\nu}= \frac{8\pi G}{ c^{4}} \langle\Psi
|T_{\mu\nu}(g_{\alpha\beta})|\Psi \rangle
\end{equation}
and this is a very difficult task. Solving these equations
requires to know an explicit expression for the expectation values
of the quantum stress-energy tensor for a large family of metrics,
necessarily including those that could potentially be the solution
of the semiclassical equations. Moreover, the quantities
$\langle\Psi |T_{\mu\nu}(g_{\alpha\beta})|\Psi \rangle$ depend
also on the quantum state of the matter $|\Psi \rangle$, and the
way one fixes this  dependence is a non-trivial issue.

Due to the static character of the classical Schwarzschild
spacetime, a natural state to consider in the fixed background
approximation is the one defined with respect to the Schwarzschild
time ``t''. This is the so-called Boulware vacuum state
$|B\rangle$ \cite{boulware} and is the state that most closely
reproduces the familiar notion of Minkowski vacuum at infinity.
The evaluation of the expectation values $\langle B |T_{\mu\nu}|B
\rangle$ shows that they vanish at infinity but are, for a
free-falling observer, highly divergent when $r \to r_S\equiv
2GM/c^2$ \cite{chrisful}. This ``drawback'' of the state
$|B\rangle$ has a natural interpretation. It describes the (vacuum
polarization) exterior to a static star but  it cannot describe
the exterior of a collapsing body producing a black hole. To
eliminate this divergence we need to replace $|B\rangle$ with
another state. However, the consequence of cancelling the
divergence at the horizon is the emergence of a non-vanishing
thermal flux (with a particular temperature) in the late-time
asymptotic future. This flux is associated to the Hawking emission
(see \cite{fabbrinavarro05} for a discussion based on the
equivalence principle).

However, the fact that $\langle B |T_{\mu\nu}|B \rangle$ gets
divergent at the horizon means that the semiclassical  corrections
to the Schwarzschild metric would be  very large when approaching
the surface $r=r_S$. This opens the question about the geometry of
the spacetime in the vicinity of $r_S$ once backreaction effects
are properly included. One usually disregards this question
arguing, as we have already stressed, that the Boulware state is
not the appropriate one to describe a collapsing star. However,
one could expect that the type of divergence of $\langle B
|T_{\mu\nu}|B \rangle$ is, in some way, related to the late-time
radiation properties of the black hole. This is indeed what
happens in the fixed background approximation. Moreover, in
addition to this there is another motivation to study this
problem. It appears in an apparently different scenario, namely in
braneworld models in Anti-de Sitter space. There is an intriguing
holographic relation between the quest of static black hole
solutions in AdS branworlds and the problem of finding consistent
solutions to the Einstein semiclassical equations in the Boulware
state. This is so according to an extension, applied to the
Randall-Sundrum model \cite{ransun}, of the Maldacena AdS/CFT
duality \cite{malda}. One expects that ``4D black holes localized
on the brane found by solving the classical bulk equations in
$AdS(5)$ are quantum corrected black holes (in the Boulware state)
and not classical ones'' \cite{efk, ta}. Significative evidence
for this conjecture has recently been given in \cite{abf} through
a numerical computation of $\langle B |T_{\mu\nu}|B \rangle$ at
large $r$. More details can be found in the contribution
\cite{fabbri05} in this Conference.

To solve the backreaction equations requires an exact analytical
expression of $\langle B |T_{\mu\nu}|B \rangle$ for a generic
geometry. Since no such expression exists in the literature (for
analytic approximations see \cite{ahs, frolovnovikov}) we resort
to the so-called $s$-wave approximation. This means to assume
spherical symmetry for the background and keep only the $s$-wave
contribution of the matter prior to quantization. In this
situation a generic expression for $\langle B |T_{\mu\nu}|B
\rangle$ can be worked out  \cite{fabbrifaresenavarro03, icp05}
generalizing the well-known results in two-dimensional spacetimes
\cite{daviesfullingunruh76, birreldavies}. In section 2 we briefly
present this approximation scheme and, in section 3, we focus on
the Boulware vacuum. Our results are exposed in section 4. We pay
attention to describe how the non-perturbative solution to the
backreaction equations prevents the formation of an event horizon
and how a singularity emerges after a bouncing point for the
radial function. Finally, in section 5 we discuss on these
results.

\section{$S$-wave approximation and semiclassical equations}
The Hilbert-Einstein action coupled to a massless scalar field \be
S^{(4)}=\frac{c^3}{16\pi G}\int d^4x\sqrt{-{g^{(4)}}}R^{(4)}
-\frac{1}{8\pi}\int d^4x\sqrt{-g^{(4)}}(\nabla f)^2
  \ee
can be reduced, under the assumption of spherical symmetry
$ds_{(4)}^2=ds_{(2)}^2+r^2 d\Omega^2$ and keeping only  the
$s$-wave component of the expansion for the matter field \be
f=f(x^a)\equiv \frac{f_{l=0}}{r}Y_{00} , \ee
 to the following effective two-dimensional theory
\be S=\frac{c^3}{4G}\int d^2x\sqrt{-g}\left [r^2R+2\left(1+|\nabla
r|^2\right)  -\frac{1}{2}r^2(\nabla f)^2 \right ]\ . \ee

The equations of motion obtained by varying directly the action
$S$ are \bea \frac{c^4}{4G}\left [-2r\nabla_{a}\nabla_{b}r
+g_{ab}\left( 2r \nabla^2 r -1+|\nabla r|^2\right)
\right]&=& T_{ab} \ , \nonumber \\
\frac{c^3}{G}\left [r R-2\nabla^2 r \right ]&=&
-\frac{2}{\sqrt{-g}}\frac{\delta S_m}{\delta r} \ ,
\label{eq:eom-Pol-2}\end{eqnarray} where $T_{ab}$ is the
two-dimensional stress-energy tensor and $S_m$ the matter sector
of the action  \be T_{ab} \equiv -\frac{2c}{\sqrt{-g}}\frac{\delta
S_m}{\delta g^{ab}} \ . \ee The quantities in the right hand side
of Eqs. (\ref{eq:eom-Pol-2}) are related to the four-dimensional
stress-energy tensor as follows: $T^{(4)}_{ab}=\frac{T_{ab}}{4\pi
r^2}, T^{(4)}_{\theta\theta}=
{T^{(4)}_{\varphi\varphi}}{\sin^{-2}\theta}=-\frac{rc}{8\pi\sqrt{-g^{(2)}}}\frac{\delta
S_m}{\delta r}$. To construct the semiclassical theory we need an
expression for the expectation values $\langle T_{ab}\rangle$ and
$\langle \frac{\delta S_m}{\delta r}\rangle$. Remarkably this can
be done in a very simple  way  by working in the conformal gauge
($ds^2=-e^{2\rho}dx^+dx^-$) for the two-dimensional part of the
metric. The following natural conditions:
\begin{enumerate}
\item the  four-dimensional covariant conservation law, which in
our two-dimensional language reads  \be \nabla^a \langle
T_{ab}\rangle=\nabla_b r \frac{1}{\sqrt{-g}}\langle\frac{\delta
S_m}{\delta r}\rangle
 \ , \ee
 \item at an arbitrary  point $X$ of the spacetime manifold the
expectation values of the quantum stress-energy tensor $\langle
T_{\pm\pm}(x^{\pm}(X))\rangle$ reduce to the normal ordering ones
$\langle :T_{\pm\pm}(x^{\pm}(X)):\rangle$ when using
 a locally inertial frame $\xi^{\alpha}_X$ based on that point  \be
\langle T_{\pm\pm}(\xi^{\alpha}_X(X))\rangle = \langle
:T_{\pm\pm}(\xi^{\alpha}_X(X)):\rangle , \ee
\end{enumerate}
related to  basic ingredients of general relativity: i) covariance
and ii) equivalence principle, are enough to provide an expression
for $\langle \Psi |T_{\pm\pm}|\Psi\rangle$, $\langle\Psi
|T_{+-}|\Psi\rangle$ and $\langle\Psi | \frac{\delta S_m}{\delta
r}|\Psi\rangle$ (here we write explicitly the quantum state
 $|\Psi\rangle$)\bea \label{uno}\langle\Psi |T_{\pm\pm}(x^\pm)|\Psi \rangle
&=&-\frac{\hbar}{12\pi}(\partial_\pm\rho\partial_\pm\rho-\partial^2_\pm\rho)+\frac{\hbar}{2\pi}
(\partial_\pm\rho\partial_\pm\phi+\rho(\partial_\pm\phi)^2)
\nonumber \\ &+&\langle\Psi
|:T_{\pm\pm}(x^\pm):|\Psi \rangle \ , \ \ \ \ \ \\
\label{due} \langle\Psi |T_{+-}(x^\pm)|\Psi \rangle &=&
-\frac{\hbar}{12\pi}(\partial_+\partial_-\rho
+3\partial_+\phi\partial_-\phi-3\partial_+\partial_-\phi) \ , \\
\label{tre} \langle\Psi|\frac{\delta
S_m}{\delta\phi}|\Psi\rangle&=&-\frac{\hbar}{2\pi}(\partial_+\partial_-\rho+\partial_+\rho\partial_-\phi+\partial_
-\rho\partial_+\phi \nonumber +2\rho\partial_+\partial_-\phi) \
\\ &+&\langle\Psi|\frac{\delta
S_m}{\delta\phi}|\Psi\rangle_{\rho=0} \ , \eea where we have
introduced, not to break with tradition in this area, the dilaton
field $\phi$ defined as $r=r_0 e^{-\phi}$ ($r_0$ is an arbitrary
length scale). The dependence on the quantum state is all
contained in the three functions
$\langle\Psi|:T_{\pm\pm}:|\Psi\rangle \equiv
\langle\Psi|T_{\pm\pm}|\Psi\rangle_{\rho=0} $  and $ \langle
\Psi|\frac{\delta S_m}{\delta \phi}|\Psi\rangle _{\rho=0}$. These
functions are not independent and verify the following relations
\be \label{conslawflatspace}
\partial_{\mp}\langle\Psi|:T_{\pm\pm}:|\Psi\rangle + \partial_{\pm}  \phi \langle \Psi|\frac{\delta
S_m}{\delta \phi}|\Psi \rangle _{\rho=0}- \frac{\hbar}{4\pi
}\partial_{\pm}(\partial_+\phi \partial_-\phi -
\partial_+\partial_-\phi) =0\ . \ee
We note that the non-vanishing of $\langle T_{+-}\rangle$ implies
the existence of a trace anomaly, absent in the classical theory.
The specific value of  $\langle T_{+-}\rangle$ is related to the
anomalous transformation law of
$\langle\Psi|:T_{\pm\pm}:|\Psi\rangle$ under a conformal rescaling
of coordinates. Moreover, the expressions (\ref{uno}) reduce, when
the value of $\phi$ is fixed, to the well-known expressions of a
conformal scalar field in two-dimensions
\cite{daviesfullingunruh76, birreldavies, giddings04,
strominger05, thorlacious95}: $\langle\Psi|T_{\pm\pm}|\Psi\rangle=
\langle\Psi|T_{\pm\pm}|\Psi\rangle_{\rho=0}
-\frac{\hbar}{12\pi}(\partial_\pm\rho\partial_\pm\rho-\partial^2_\pm\rho)$,
or equivalently, $\langle\Psi|T_{\pm\pm}|\Psi\rangle=
-\frac{\hbar}{12\pi}(\partial_\pm\rho\partial_\pm\rho-\partial^2_\pm\rho+
t_{\pm})$, with the identification $-\frac{\hbar}{12\pi
}t_{\pm}=\langle\Psi|:T_{\pm\pm}:|\Psi\rangle$.

 \section{Semiclassical equations in the Boulware state}
Since we are interested in the state which more naturally mimics
the Minkowski vacuum in flat space (i.e., the Boulware state) we
shall define it by using, in quantizing the matter, the time
coordinate ``t'' respect to which the metric takes the static
form. Therefore it is natural to impose  that \be \label{condboul}
\langle B|:T_{\pm\pm}(t,x):| B\rangle =\langle B|T_{\pm\pm}(t,x)|
B\rangle_{\rho=0}= 0\ . \ee This definition assumes that the
semiclassical background metric is  static: $\rho=\rho(x)$ and
$\phi = \phi(x)$, where $x=(x^+ - x^-)/2$. A straightforward
consequence of the above equation is that the expectation value
$\langle \Psi|\frac{\delta S_m}{\delta \phi}|\Psi \rangle
_{\rho=0}$  can be determined immediately \be
\label{condphi}\langle B|\frac{\delta S_m}{\delta \phi}| B \rangle
_{\rho=0}=-\frac{\hbar}{16 \pi}
\frac{(\phi_{x}^2-\phi_{xx})_{x}}{\phi_x} \ , \ee where the
subindex $x$ means  derivative with respect to the coordinate $x$.
Therefore we have all ingredients to write down the semiclassical
Einstein equations in the $s$-wave approximation in the Boulware
vacuum. Fixing, for simplicity, the value of the parameter $r_0$
as $r_0 \equiv \sqrt{\lambda} =\sqrt{\frac{\hbar G}{12\pi
c^3}}\equiv \sqrt{\frac{l_{Planck}^2}{12\pi}}$, the final result
is
\begin{eqnarray}
\phi_{xx}-\phi_x^2-2\rho_x\phi_x&=&e^{2\phi}\left[\rho_{xx}-\rho_x^2+6\rho_x\phi_x+6\rho\phi_x^2\right] \label{eq:cg-1}\ , \\
\phi_{xx}-2\phi_x^2+\frac{e^{2(\phi+\rho )}}{\lambda}&=& e^{2\phi}\left[\rho_{xx}-3(\phi_{xx}-\phi_x^2)\right]  \label{eq:cg-2}\ , \\
\phi_{xx}-\phi_x^2-\rho_{xx}&=&e^{2\phi}\left[3\rho_{xx}+6\rho_x\phi_x+6\rho\phi_{xx}+\frac{3}{2}\frac{(\phi_{xx}-\phi_x^2)_x}{\phi_x}
\right]. \ \ \ \label{eq:cg-3}
\end{eqnarray}
When the right hand side is zero, the solution describes  the
Schwarzschild geometry \bea \label{classicalsolutions2} \rho&=&
\frac{1}{2}\ln(1-\frac{2GM}{c^2
r})  \\
r^*\equiv x &=& r + \frac{2GM}{c^2}\ln(1-\frac{2GM}{c^2r})
\label{classicalsolutions21} \ , \eea with mass $M$. In the
quantum theory the solution deviates from the above expressions
but we shall impose that, for large $r$, the semiclassical
solution approaches the classical one.

\section{Exact Semiclassical solutions}

Our task is to investigate  how the classical relations
(\ref{classicalsolutions2}--\ref{classicalsolutions21}) for
$\rho=\rho(r)$ and $ r=r(x)$ are modified by pure vacuum
polarization  effects. The main properties of the classical
solutions are \begin{itemize} \item The function $r=r(\rho)$ is
monotonic, with \be \frac{dr}{d\rho} > 0 \ , \ee reaching $\rho =
-\infty$ at the finite value $r=r_S$. \item The function $r(x)$ is
monotonic, with \be \frac{dr}{dx} > 0 \ , \ee reaching $x=-\infty$
at $r=r_S$.
\end{itemize}

The numerical solution for the semiclassical equations violates
the above properties and unravels the following features (the
details of the integration can be found in \cite{ffnoh}):
\begin{itemize} \item {\bf Existence of a
bouncing point for the radial function.} The function $r=r(\rho)$
is very similar to the classical solution till very close to
$r_S$. Just before $r_S$ the function $r=r(\rho)$ has a minimum.
This bouncing point $r= r_B > r_S $ for the radial function \be
 \frac{dr}{d\rho}(r_B) = 0 \ , \ee
 takes
place at a very small but non-zero value of $g_{00}$. In the
vicinity of this point the metric can be approximated by \be
\label{approxbounce}ds^2 \approx -e^{2\rho(r_B)}c^2dt^2 +
\alpha\frac{dr^2}{(1-\frac{r_B}{r})} + r^2d\Omega^2
 \ , \ee
 where $\alpha$ is a numerical constant.

 \item{\bf Existence of a branching point for the radial
 function}. The function $r=r(x)$ is very similar to the classical
solution till very close to $r_S$. Just before $r_S$ the function
$r=r(x)$ has a minimum, corresponding to the bouncing point
$r=r_B$ described above. \be
 \frac{dr}{dx}(r_B) = 0 \ . \ee
Moreover, just after it we encounter a minimum for the
``tortoise'' coordinate $x$ \be \frac{dx}{dr}(x_M)=0 \ , \ee at a
finite value $x=x_M$. The dependence of the radial function around
this point is \be\label{rbranching} r\approx
r_M-\beta(x-x_M)^{2/3}\ee where $\beta$ is a positive numerical
constant. The radial function has a branching point at $x=x_M$,
which turns out to be the minimal possible value for the
coordinate $x$. Although all components of the metric are finite
at $x=x_M$  the branching (\ref{rbranching}) generates a curvature
singularity for the metric.

\end{itemize}

\section{Final comments}

The most important result of the investigation presented in this
contribution is the emergence of a bouncing point for the radial
function in the static quantum corrected Schwarzschild geometry.
It would be interesting to see if this feature is also reproduced
in the static solutions in braneworld models in $5$-dimensional
Anti-de Sitter space.

More explicitly, the classical solution for the tortoise
coordinate \be r^* \equiv x = r + r_S\ln \frac{|r-r_S|}{r_S}  \ee
is modified, in the vicinity of $r_B \approx r_S$, by \be x
\approx x_B + e^{-\rho(r_B)}\sqrt{\frac{r_B}{\alpha}(r-r_B)} \ .
\ee Naive use of the standard formulaes to derive the emitted
radiation would convert the thermal Hawking luminosity $L \propto
T_H^2$, where $T_H$ is the Hawking temperature, into \be L \propto
\frac{1}{(x^- - x^-_B)^2} \ , \ee which is unbounded when the
retarded time $x^-_B$ corresponding to the bounce point is
reached. This seems to indicate that backreaction effects can
produce significative  changes to the standard view of the
evaporation process.

\end{document}